\documentclass{aa}  
\usepackage{graphicx}
\usepackage{textgreek} 
\usepackage{txfonts}
\usepackage{multirow}
\usepackage{dcolumn}
\usepackage[usenames,dvipsnames]{color}
\begin{document}

   \title{Diffuse radio emission in the complex merging galaxy cluster Abell\;2069}

   \authorrunning{A. Drabent et al.}

   \titlerunning{Diffuse radio emission in the complex merging galaxy cluster A2069}

   \author{A. Drabent\inst{1}
          \and
          M. Hoeft\inst{1}
	  \and
	  R. F. Pizzo\inst{2}
	  \and
	  A. Bonafede\inst{3}
	  \and
	  R. J. van Weeren\inst{4}
	  \and
	  U. Klein\inst{5}
          }

   \institute{Th\"uringer Landessternwarte (TLS), Sternwarte 5, D-07778 Tautenburg, Germany \\
              \email{alex@tls-tautenburg.de}
         \and
	     ASTRON -- Netherlands Institute for Radio Astronomy, Oude Hoogeveensedijk 4, 7991 PD Dwingeloo, The Netherlands
	 \and
	     Hamburger Sternwarte, Gojenbergsweg 112, D-21029 Hamburg, Germany
	 \and
	     Harvard-Smithsonian Center for Astrophysics, 60 Garden Street, Cambridge, MA 02138, USA
	 \and
             Argelander Institut f\"ur Radioastronomie, Universit\"at Bonn, Auf dem Hügel 71, D-53121 Bonn, Germany
             }

   \date{Received December 18, 2014; accepted December 18, 2014}

  \abstract
   {Galaxy clusters with signs for a recent merger show in many cases extended diffuse radio features. This emission originates from relativistic electrons which suffer synchrotron losses due to the intra-cluster magnetic field. The mechanisms of the particle acceleration and the properties of the magnetic field are still poorly understood.}
   {We search for diffuse radio emission in galaxy clusters. Here, we study the complex galaxy cluster Abell\,2069, for which X-ray observations indicate a recent merger.}
   {We investigate the cluster's radio continuum emission by deep Westerbork Synthesis Radio Telescope (WSRT) observations at 346 MHz and a Giant Metrewave Radio Telescope (GMRT) observation at 322\,MHz.}
   {We find an extended diffuse radio feature roughly coinciding with the main component of the cluster. We classify this emission as a radio halo and estimate its lower limit flux density to $25\pm9$\,mJy. Moreover, we find a second extended diffuse source located at the cluster's companion and estimate its flux density to $15\pm2$\,mJy. We speculate that this is a small halo or a mini-halo. If true, this cluster is the first example of a double-halo in a single galaxy cluster.}
   {}


   \maketitle
\section{Introduction}
Diffuse radio emission has been found in about one hundred galaxy clusters.
Depending on location in the cluster, morphology, and spectral properties one distinguishes radio relics and radio halos \citep[see][for a review]{2012A&ARv..20...54F}.
There is strong evidence that both relics and halos are related to a recent merger event in galaxy clusters \citep{2010ApJ...721L..82C,2013ApJ...777..141C,2012PASJ...64...49A}.
Scenarios for the formation of radio halos can be subdivided into `hadronic models' where secondary electrons, generated by hadronic collisions of relativistic protons with thermal nuclei, cause the radio emission \citep[e.g.][]{1980ApJ...239L..93D,1999APh....12..169B} and `turbulence models' in which electrons are re-accelerated by the turbulence in the intra-cluster medium in the aftermath of the cluster merger \citep{2001MNRAS.320..365B,2001ApJ...557..560P}.
The `hadronic model' is disfavored by several observational evidences, e.g. the non-detection of \textgamma-ray emission in nearby galaxy clusters \citep{2009A&A...508..599B,2010ApJ...717L..71A,2012MNRAS.426..956B,2014ApJ...787...18A,2014MNRAS.440..663Z}, the discovery of ultra-steep spectrum halos \citep{2008Natur.455..944B} and the radial profile of some giant radio halos \citep[e.g.][]{2014MNRAS.438..124Z}.
Moreover, some merging clusters do show halo emission while other merging systems do not.
This observational fact may provide further constraints on the origin of radio halos \citep{2010ApJ...721L..82C,2013ApJ...777..141C,2011MNRAS.417L...1R}.
For a comprehensive theoretical overview of the scenarios of the formation of radio halos, see the review by \citet{2014IJMPD..2330007B}.

In some galaxy clusters without a recent merger but with a cool core also small radio halos have been found \citep[][for a recent compilation]{2012A&ARv..20...54F,2014ApJ...781....9G}.
This emission is referred to as `mini-halo'.
Cold fronts, as found in Abell\,2069 \citep{2009ApJ...704.1349O}, have been observed in some clusters possessing a mini-halo.
It has been suggested that the cold fronts are related to gas sloshing, which in turn generates turbulence in the cluster core, which then re-accelerates relativistic electrons \citep{2002A&A...386..456G,2004A&A...417....1G,2008ApJ...675L...9M,2013ApJ...762...78Z}.
Hadronic scenarios as described above have also been suggested for mini-halos \citep{2004A&A...413...17P,2010ApJ...722..737K,2013MNRAS.428..599F} and are studied in detailed simulations \citep{2014arXiv1403.6743Z}.
For mini-halos the \textgamma-ray emission is significantly lower than for giant halos.
Hence, current \textgamma-ray observations do not achieve enough sensitivity to provide further constraints on the hadronic origin of mini-halos \citep{2014IJMPD..2330007B}.
The giant radio halo found in the cool-core cluster CL1821+643 \citep{2014arXiv1407.4801B} could be an example of a transitional stage between radio halos and mini-halos.
Similarly, a mini-halo in a merging system may also provide evidence that giant halos and mini-halos are closely related to each other.

In this paper we present radio observations of the Abell\,2069 galaxy cluster complex with the WSRT at 346\,MHz and the GMRT at 322\,MHz.
We assume a \textLambda CDM cosmology with $H_0 = 71$\,km\,s$^{-1}$\,Mpc$^{-1}$, \textOmega$_{\text{m}}=0.27$, and \textOmega$_{\text{\textLambda}}=0.73$.
For a redshift $z = 0{.}116$ this translates to 2.076\,kpc/$^{\prime\prime}$. 

\section{Abell\,2069}
\begin{figure*}[htbp]
\centering
\includegraphics[trim=65 175 90 230,clip,scale=0.5]{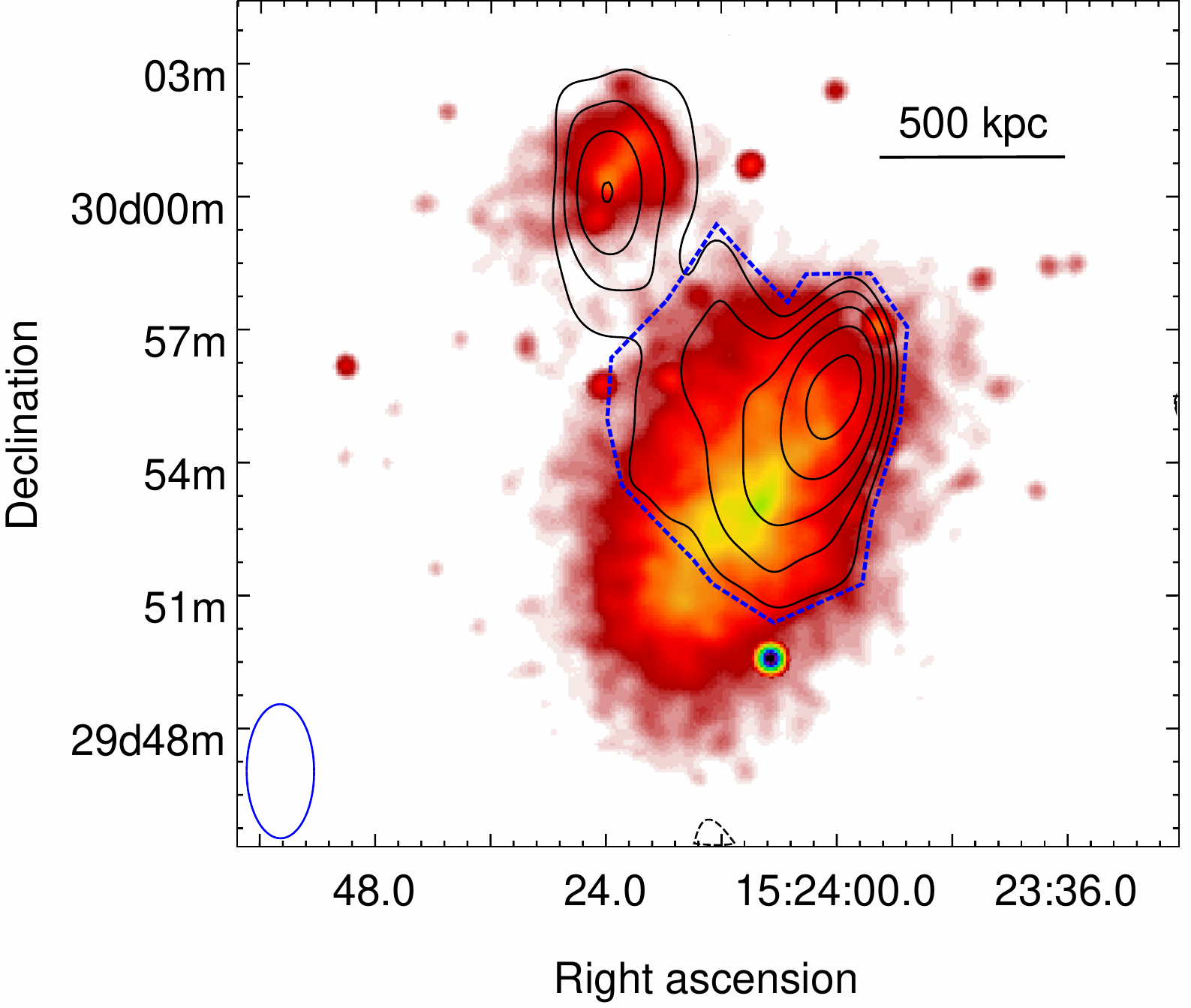}
\put(-50,70){\makebox(0,0){\textcolor{black}{\textbf{\huge{A}}}}}
\put(-137,155){\makebox(0,0){\textcolor{black}{\textbf{\Large{B}}}}}
\includegraphics[trim=65 175 90 230,clip,scale=0.5]{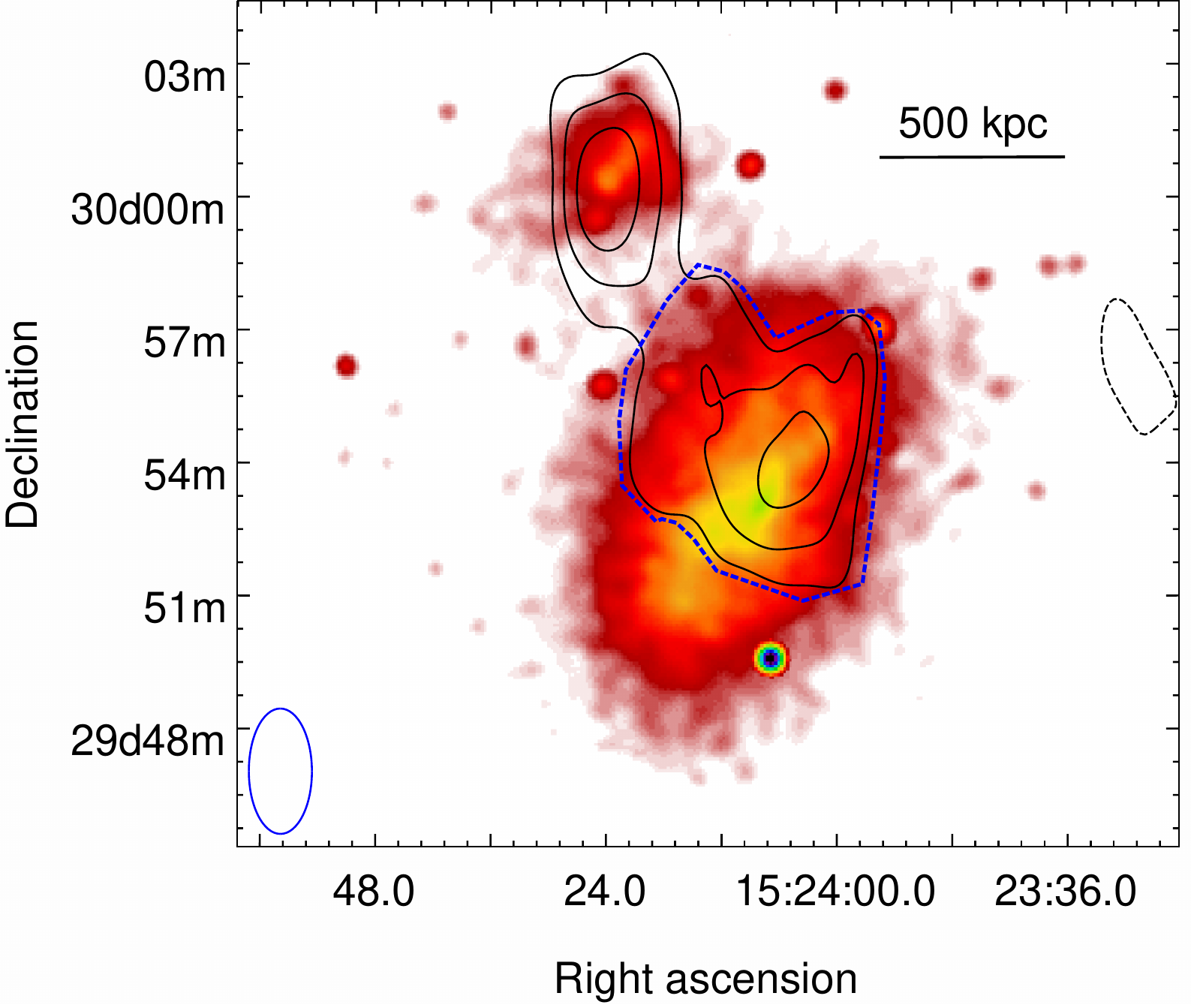}
\put(-50,70){\makebox(0,0){\textcolor{black}{\textbf{\huge{A}}}}}
\put(-137,155){\makebox(0,0){\textcolor{black}{\textbf{\Large{B}}}}}
\caption{\textbf{Left image:} \textit{Colorscale:} X-ray surface brightness at 0{.}5\,-\,7\,keV in linear scale smoothed with a 2D Gaussian of $\sigma=3^{\prime\prime}$. The observation has been taken with the Chandra ACIS-I. The cluster components are labelled with `A' and `B'. \textit{Black contours:} WSRT 346\,MHz image after subtracting compact sources. The noise level is $\sigma_{\text{rms}}\approx 1{.}0$\,mJy/beam. Contour levels are drawn at $\left[-3{.}0{,}\,3{.}0{,}\,4{.}2\,{,}\,6{.}0{,}\,8{.}5{,}\,12{.}0\right]$\,mJy. The negative contour is drawn with a dashed line. \textit{Blue dashed line:} Region where the radio flux density related to the main cluster component A has been measured. The skymodel for subtraction is based on the measured flux densities in the GMRT image at 322 MHz, listed in Tab.\;\ref{tab:A2069}. The shape of the restoring beam is $182^{\prime\prime}\times91^{\prime\prime}$ with a position angle of $0^{\circ}$. \textbf{Right image:} Image obtained using 
a more aggressive subtraction procedure. The shape of the restoring beam is $170^{\prime\prime}\times85^{\prime\prime}$ with a position angle of $0^{\circ}$.}
\label{fig:chandra}
\end{figure*}

The cluster Abell\,2069 is a complex galaxy cluster of richness class 2 at a redshift of $z=0{.}116$ \citep{1999ApJS..125...35S}.
Together with five other clusters it belongs to the so-called `A2069-supercluster' ($z=0{.}11$) \citep{1997A&AS..123..119E}.
The Corona Borealis supercluster \citep{1997ApJS..111....1S,1997ApJ...487..512S} is located at the foreground ($z=0{.}07$).
Abell\,2069 is a merger system with two bright elliptical galaxies \citep{1982ApJ...255L..17G} in the center with a projected distance of about 55\,kpc.
In X-ray observations two components, in the following denoted A and B, are visible and $\approx$\,1\,Mpc separated from each other (see Fig.\;\ref{fig:chandra}).
The cluster component B has a peculiar velocity of 485\,km\,s$^{-1}$ with respect to the mean cluster redshift and shows a cold front \citep{2009ApJ...704.1349O}.
Properties of Abell\,2069 are summarized in Tab.\,\ref{tab:A2069}.
\begin{table}[htbp]
\caption{Cluster properties and flux densities of compact sources in Abell\,2069 as measured with the GMRT at 322\,MHz.}
\begin{center}
\footnotesize{
\begin{tabular}{lD{X}{\,\pm\,}{-1}}
\hline\hline
RA (J2000)  [h m s]                                               & \multicolumn{1}{c}{15 24 09.8}                 \\
DEC (J2000) [$^\circ$ $^\prime$ $^{\prime\prime}$]                & \multicolumn{1}{c}{+29 55 16}                  \\
$z$                                                               & \multicolumn{1}{D{.}{.}{-1}}{0.116^{\text{a}}} \\
$L_{\text{X}}$ (0{.}1\,-\,2{.}4\,keV)\;[$10^{44}$\,erg\,s$^{-1}$] & \multicolumn{1}{D{.}{.}{-1}}{4.55^{\text{b}}}  \\[1.5ex]
flux densities of compact sources [mJy]:                          &                                                \\
\;\;\;\;A1$^{\star}$                                              & 1{.}6X0{.}3                                    \\
\;\;\;\;A2                                                        & 1{.}0X0{.}3                                    \\
\;\;\;\;B                                                         & 3{.}8X0{.}6                                    \\
\;\;\;\;C$^{\dag}$                                                & 13X2                                           \\
\;\;\;\;D1                                                        & 0{.}7X0{.}1                                    \\
\;\;\;\;D2                                                        & 0{.}6X0{.}1                                    \\
\;\;\;\;E1$^{\star}$                                              & 5{.}8X0{.}8                                    \\
\;\;\;\;E2                                                        & 2{.}5X0{.}4                                    \\
\;\;\;\;F$^{\star}$                                               & 5{.}2X0{.}7                                    \\
\;\;\;\;G1$^{\star}$                                              & 4{.}6X0{.}8                                    \\
\;\;\;\;G2$^{\star}$                                              & 46\phantom{.0}X5                               \\
\;\;\;\;G3$^{\star}$                                              & 1{.}0X0{.}3                                    \\
\;\;\;\;I$^{\dag}$                                                & 14\phantom{.0}X2                               \\
\;\;\;\;J1                                                        & 0{.}8X0{.}1                                    \\
\;\;\;\;J2                                                        & 1{.}6X0{.}4                                    \\
\;\;\;\;K$^{\star}$                                               & 1{.}7X0{.}4                                    \\
\hline\hline
\end{tabular}}
\end{center}
\label{tab:A2069}
\footnotesize{
{\textbf{Notes.} All sources marked with $^{\dag}$ are cataloged in NVSS and all sources which are marginally visible in the NVSS postage stamp images are marked with $^{\star}$.}
\\{\textbf{References.} $^{\text{a}}$\citet{1999ApJS..125...35S}, $^{\text{b}}$\citet{2013ApJ...779..189F}}}
\end{table}
Searching for radio halos \citet{2013ApJ...779..189F} studied the cluster with the Green-Bank-Telescope (GBT) at 1{.}4\,GHz and discovered extended emission in Abell\,2069.
However, due to the very large beam size of the GBT conclusions about the morphology or the internal structure of the emission are limited.
Furthermore, unidentified compact sources may contribute to their flux density estimate.

\section{Observations and data reduction}
The galaxy cluster Abell\,2069 has been observed for 36 hours in P-band with the Westerbork Synthesis Radio Telescope (WSRT).
The observation was performed during three different nights with a duration of 12 hours each.
We observed in the frequency range from 311 to 381\,MHz, with a central frequency of 346\,MHz and a spectral resolution of 78.125\,kHz.
In order to maximize the uv-coverage we observed each night with another array configuration.
The WSRT allows to move four antennas\footnote{www.astron.nl/radio-observatory/astronomers/wsrt-guide-observations/3-telescope-parameters-and-array-configuration}.
Most adjacent antennas have a distance of 144\,m.
With a shortest baseline of a length of 36\,m the instrument is sensitive for large-scale structures up to $\sim40^{\prime}$.
This observation was part of the WSRT legacy project in which ten clusters were observed (Drabent et al., in prep.).

In addition, we have observed Abell\,2069 with the Giant Metrewave Radio Telescope (GMRT) for 4{.}8 hours.
Here we have observed in the frequency range from 306 to 339\,MHz, with a central frequency of 322\,MHz and a spectral resolution of 130.208\,kHz.
Details of all observations are listed in Table \ref{tab:observations}.
\begin{table*}[htbp]
\centering
\caption{Observational details. The label of the WSRT configuration is named according to the distance of antenna 9 and A in meters}
\footnotesize{
\begin{tabular}[]{cccccccccc}
\hline\hline
telescope & pointing center & frequency  & exposure time & date               & configuration       \\ 
          & RA/DEC (J2000)  & [MHz]      & [h]           &                    &                     \\ 
\hline
WSRT      & 15h23m57.9s +29$^\circ$53$^\prime$26.0$^{\prime\prime}$ & 311-381\,MHz & 12  & 26/27-April-2012 & 36M \\ 
          &                                                         &              &     & 09/10-May-2012   & 60M \\ 
          &                                                         &              &     & 16/17-May-2012   & 84M \\ 
GMRT      & 15h24m08.0s +29$^\circ$52$^\prime$55.0$^{\prime\prime}$ & 306-339\,MHz & 4.8 & 30-May-2014      & --  \\
\hline
\end{tabular}
\label{tab:observations}}
\end{table*}
3C147 (WSRT) and 3C286 (GMRT) have been observed for 15 minutes before and after the target observation for flux density calibration.

We performed RFI mitigation using the AOFlagger \citep{2010ascl.soft10017O} and optimized the default strategy for the WSRT as well as for the GMRT P-band observations.
Unfortunately, in the GMRT observation all baselines shorter than about 1\,500 meters are heavily affected by RFI.
A total of 30\% of the WSRT data and 40\% of the GMRT data has been flagged.

\begin{figure*}[htbp]
\centering
\makebox{\includegraphics[trim=195 292 225 275,clip,scale=1.0]{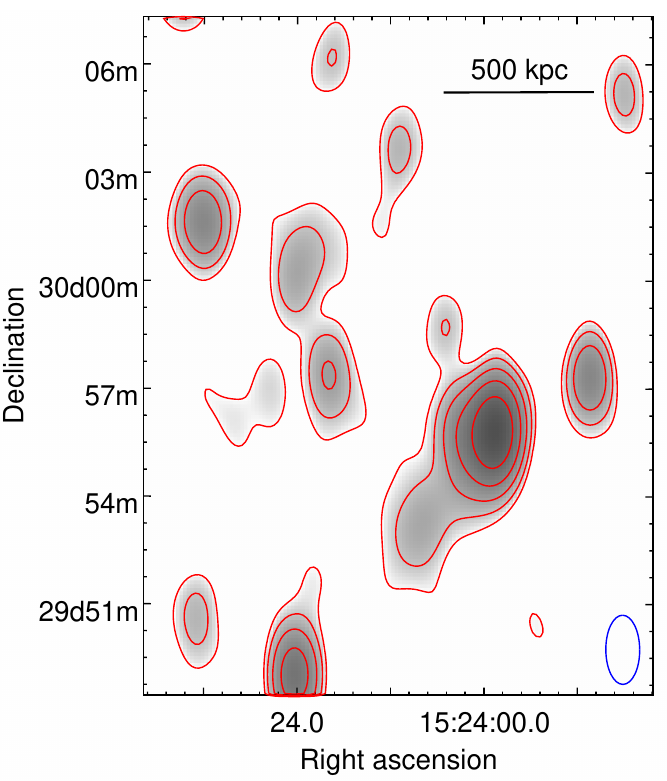}}
\put(-70,73){\makebox(0,0){\textcolor{green}{\textbf{HA}}}}
\put(-50,101){\makebox(0,0){\textcolor{green}{\textbf{G}}}}
\put(-105,148){\makebox(0,0){\textcolor{green}{\textbf{HB}}}}
\includegraphics[trim=195 292 225 275,clip,scale=1.0]{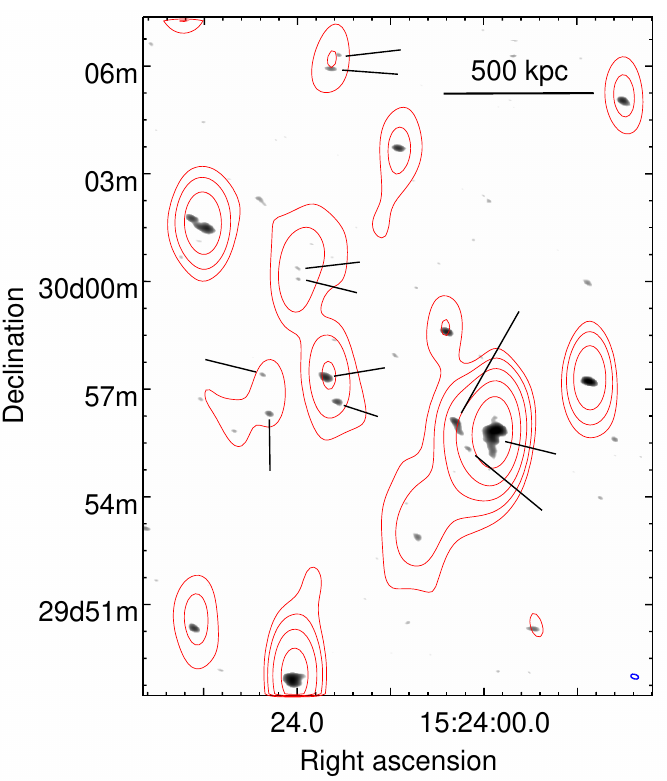}
\put(-70,213){\makebox(0,0){\textcolor{Brown}{\scriptsize{\textbf{A1}}}}}
\put(-72,203){\makebox(0,0){\textcolor{Brown}{\scriptsize{\textbf{A2}}}}}
\put(-65,183){\makebox(0,0){\textcolor{Brown}{\small{\textbf{B}}}}}
\put(-119,161){\makebox(0,0){\textcolor{Brown}{\small{\textbf{C}}}}}
\put(-82,151){\makebox(0,0){\textcolor{Brown}{\scriptsize{\textbf{D1}}}}}
\put(-84,141){\makebox(0,0){\textcolor{Brown}{\scriptsize{\textbf{D2}}}}}
\put(-75,120){\makebox(0,0){\textcolor{Brown}{\scriptsize{\textbf{E1}}}}}
\put(-78,104){\makebox(0,0){\textcolor{Brown}{\scriptsize{\textbf{E2}}}}}
\put(-55,131){\makebox(0,0){\textcolor{Brown}{\small{\textbf{F}}}}}
\put(-37,138){\makebox(0,0){\textcolor{Brown}{\scriptsize{\textbf{G1}}}}}
\put(-25,93){\makebox(0,0){\textcolor{Brown}{\scriptsize{\textbf{G2}}}}}
\put(-30,76){\makebox(0,0){\textcolor{Brown}{\scriptsize{\textbf{G3}}}}}
\put(-10,116){\makebox(0,0){\textcolor{Brown}{\small{\textbf{I}}}}}
\put(-138,122){\makebox(0,0){\textcolor{Brown}{\scriptsize{\textbf{J1}}}}}
\put(-115,85){\makebox(0,0){\textcolor{Brown}{\scriptsize{\textbf{J2}}}}}
\put(-72,76){\makebox(0,0){\textcolor{Brown}{\scriptsize{\textbf{K}}}}}
\caption{\textbf{Left image:} WSRT 346 MHz radio map imaged with Briggs weighting (\texttt{robust}\,=\,0.0 in CASA). The noise level is $\sigma_{\text{rms}}\approx 0{.}5$\,mJy/beam. Contour levels are drawn at $\left[1{.}5{,}\,3{.}0{,}\,9{.}0{,}\,18{,}\,36\right]$\,mJy. There is no negative flux density below $-1{.}5$\,mJy within the depicted area. The shape of the restoring beam is $108^{\prime\prime}\times52^{\prime\prime}$ with a position angle of $0^{\circ}$. \textbf{Right image:} \textit{Red contours:} See the left image. \textit{Greyscale:} GMRT 322\,MHz radio map imaged with Briggs weighting (\texttt{robust}\,=\,0.0 in CASA). The map is logarithmically scaled. Flux density cutoff is set to $3\times\sigma_{\text{rms}}$ with $\sigma_{\text{rms}}\approx0{.}2$\,mJy/beam. The shape of the restoring beam (denoted with the blue ellipse) is $14^{\prime\prime}\times7^{\prime\prime}$ with a position angle of $73^{\circ}$.}
\label{fig:A2069}
\end{figure*}
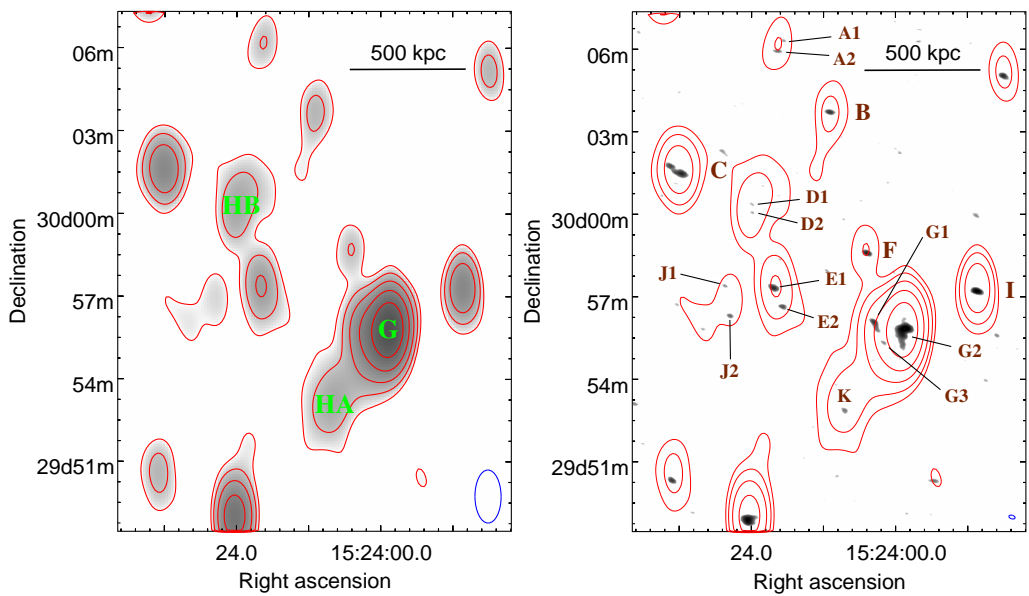

Calibration, imaging and self-calibration were carried out with the Common Astronomy Software Applications\footnote{http://casa.nrao.edu/} (CASA) package.
For the WSRT we have corrected for amplitude variations based on the system temperature information recorded during the observations.
The observations of 3C147 and 3C286 were used to correct for the bandpass and the flux density scale according to \citet{2012MNRAS.423L..30S}.
Because of the excellent phase stability of the WSRT backend a few iterations of self-calibration were sufficient to correct for remaining phase variations.
The initial phase calibration for the GMRT data was carried out with a skymodel based on positions in the FIRST catalog \citep{2012yCat.8090....0B}, using 36 sources which contribute most to the visibilities.
For self-calibration and imaging we applied Briggs weighting with a robustness parameter of 0.0 in CASA, which offers a compromise between high-resolution and the best signal-to-noise-ratio per beam.
With the final calibrated data we achieve an r.m.s. noise level of about 0.5\,mJy/beam for the WSRT images with a $108^{\prime\prime}\times52^{\prime\prime}$-beam, and 0.2\,mJy/beam for the GMRT images with a $14^{\prime\prime}\times7^{\prime\prime}$-beam.
Both radio maps obtained with this procedure are depicted in Fig.\;\ref{fig:A2069}.

Throughout our flux density measurements we assume a flux density scale uncertainty of 5\% for 3C147 and 2.5\% for 3C286 according to \citet{2012MNRAS.423L..30S}.
All errors given below take into account noise uncertainties based on the r.m.s. level of the radio maps.
Since we could not correct for system temperature variations of the GMRT we adopt an average amplitude instability of 10\%.

We complemented our analysis by using archival Chandra ACIS-I archival data (ObsID: 4965).
The actual exposure time amounts to 55{.}4 ks.
For the data reduction we used the Chandra analysis software CIAO 4.6 and CALDB 4.5.9.
The exposure-corrected image is created with the task `fluximage'. 

\section{Results and Discussion}
We have detected extended diffuse emission in both components A and B of Abell\,2069 (see Fig.\;\ref{fig:chandra}).
In the following we describe the extraction of the diffuse emission in more detail.

In Fig.\;\ref{fig:A2069} the maps of the WSRT (left image) and the GMRT (right image) observations are compared.
In both images several sources within the cluster area and its vicinity are evident.
The high-resolution GMRT image reveals that most of the sources visible in the WSRT image are compact.
For instance, the sources `B', `F' and `I' show similar flux densities in both images.
All measured flux densities of the GMRT image are presented in Tab.\,\ref{tab:A2069}.

\begin{figure}[htbp]
\centering
\includegraphics[trim= 195 290 255 275 ,clip,scale=1]{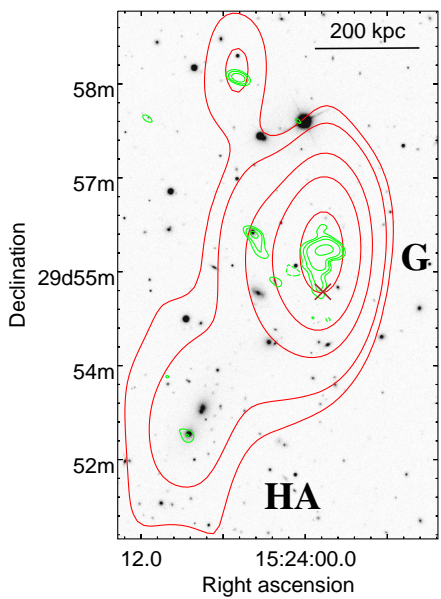}
\put(-9,130){\makebox(0,0){\textcolor{black}{\Large{\textbf{G}}}}}
\put(-55,40){\makebox(0,0){\textcolor{black}{\Large{\textbf{HA}}}}}
\caption{Zoom into the `G'/`HA' complex. \textit{Red contours:} See the red contours in Fig.\;\ref{fig:A2069}. \textit{Green contours:} 322\,MHz GMRT observation. The noise level is $\sigma_{\text{rms}}\approx 0{.}2$\,mJy/beam. Contour levels are drawn at $\left[-0{.}75{,}\,0{.}75{,}\,1{.}1{,}\,1{.}5\,{,}\,2{.}1\right]$\,mJy. \textit{Background:} SDSS-III (Data release 10) r-band image. \textit{Brown cross:} Position of the potential host galaxy producing the radio emission `G2'.}
\label{fig:SourceG}
\end{figure}

The GMRT image shows that the unresolved luminous source `G' in the WSRT image is composed of three distinct sources, where the extended source `G2' dominates.
Neither in the FIRST \citep{2012yCat.8090....0B} survey, the NVSS \citep{1998AJ....115.1693C} at 1{.}4\,GHz nor in the VLSS \citep{2007AJ....134.1245C} at 74\,MHz has a source been reported at this location.
The morphology of `G2' suggests that we see projected superimposed lobes of a radio galaxy.
Possibly, while the galaxy is moving through the intra-cluster medium both lobes could have been bent to the north.
At the southern end of the emission's tail we identified the faint galaxy SDSS\,J152358+29551 in the Sloan Digital Sky Survey (SDSS) \citep{2012ApJS..203...21A} at r-band, see the brown cross in Fig.\;\ref{fig:SourceG}.
For this galaxy a photometric redshift of $z_{\text{photo}}=0{.}100\pm0{.}042$ was determined.
Therefore, it could be a member of the galaxy cluster.

The peak positions of the sources `E1' and `E2' coincide with the foreground optical galaxies MCG+05-36-028 and WISEPC J152419+2956 (both $z=0.076$), respectively.
They thus belong to a cluster member of the Corona Borealis Supercluster \citep{1982ApJ...255L..17G}.

In the case of the sources `HA' and `HB' visible in the WSRT image no corresponding compact sources could be found in the GMRT image.
The sources `D1', `D2' and `K', which are located at the brightest cluster galaxy of the corresponding cluster component, are too faint ($\lessapprox 1$\,mJy). 
This suggests that `HA' and `HB' are extended diffuse radio features.
To better recover diffuse emission we imaged with natural weighting after subtracting compact sources.

To achieve this we used the deep WSRT observation which also offers a good coverage of the short uv-spacings.
The GMRT image serves as a skymodel to remove all visible compact sources in the area of Abell\,2069.
Since the center frequencies of both observations are slightly different we assume an overall spectral index of all sources of $\alpha=-0{.}8$. 
We then subtracted our skymodel in the uv-plane from the calibrated WSRT data and deconvolved the data again now using natural weighting.
The latter gives more emphasis on shorter baselines, so extended emission is expected to be better recovered.

The resulting radio map, depicted in the left image of Fig.\;\ref{fig:chandra}, shows extended emission in the main component of Abell\,2069 as well as a distinct feature in the subcomponent B.
The emission in A is apparently extended towards the cluster component B.
The extended radio emission of the main component A with a largest linear scale (LLS) of $\approx900$\,kpc is roughly elongated along the semi-major axis of the corresponding elliptical X-ray structure.
The peak flux is shifted to the northwest.
This might be partly caused by not entirely subtracting the source `G'.
Imaging the WSRT data with compact sources subtracted and with uniform weighting for higher resolution shows residual emission of $14\pm2$\,mJy at the position of `G'.
This indicates that the source `G' in the GMRT observation does not show all the flux density as observed with the WSRT.
We therefore increased the flux density of `G' in our skymodel by 14\,mJy and subtracted it again in the uv-plane from the calibrated WSRT data, see right image of Fig.\;\ref{fig:chandra}.
We now find a remaining LLS of $\approx$\,750\,kpc and a radio-X-ray offset of 120\,kpc (fractional offset of 0.16).

The elongated X-ray morphology of the component A and the presence of two brightest cluster galaxies indicate a recent merger.
Since the extended radio emission in A shows roughly a similar orientation as the X-ray emission with only a moderate shift in position, we classify the emission as radio halo.
Radio halos of similar size have been recently discovered in the galaxy cluster Abell\,1689 \citep{2011A&A...535A..82V} and CL\,0217+70 \citep{2011ApJ...727L..25B}.
The latter halo also shows a slight offset to the corresponding X-ray peak.

For the cluster component B we can only little constrain the morphology of the source `HB' since its size is comparable to the beam width.
Comparing the profile of `HB' to the restoring beam profile we find an extent of the source of a few ten arcseconds.
This corresponds to a LLS of the order of 50 to 100\,kpc.

In our maps both radio features in component A and B are connected by a bridge.
We speculate that this is an effect of the large beam size.

We derive a total extended flux density within the black contours of $54\pm9$\,mJy at 346\,MHz.
Note, due to flux density uncertainties of the subtracted compact sources we have additionally included 10\% of the subtracted flux density to the error estimates.
We measure a flux density of $39\pm8$\,mJy for the diffuse flux associated with the cluster main component A (in the following denoted with `HA') within the region enclosed by the blue dashed line (see left image of Fig.\;\ref{fig:chandra}).
In the map obtained with the more aggressive subtraction of `G' (see above and right image of Fig.\,\ref{fig:chandra}) we measure a flux density of $25\pm9$\,mJy.
While in the first procedure part of the flux density of `G' possibly remains in the subtracted image, with more aggressive subtraction we may also remove part of the diffuse emission of the halo.
The maps (depicted in Fig.\;\ref{fig:chandra}) thus represent an upper and a lower limit of the flux density distribution of `HA'.
For the source `HB' we measure a flux density of $15\pm2$\,mJy in both maps of Fig.\;\ref{fig:chandra}.

Our findings confirm the discovery of diffuse emission by \citet{2013ApJ...779..189F}.
After subtracting NVSS sources they recovered $28{.}8\pm 7{.}2$\,mJy of diffuse flux density at 1{.}4\,GHz.
They subtracted all sources visible in the NVSS image down to a flux density threshold of 1.35\,mJy.
Furthermore, they estimated that the recovered flux density may contain 8-13\,mJy due to faint radio sources below the threshold.
Our GMRT image at 322\,MHz corroborates the presence of several sources which are not or only marginally recognized in the NVSS (see Tab.\;\ref{tab:A2069}).
However, the total flux density of 11\,mJy in the newly discovered sources is small in comparison to 66\,mJy of all in the NVSS marginally visible sources.
Comparing our upper (54$\pm$9\,mJy) and our lower (40$\pm$11\,mJy) limit of the total diffuse flux density (`HA'+`HB') at 346\,MHz with the estimated upper ($\approx$20\,mJy) and the lower ($\approx$15\,mJy) limit at 1.4\,GHz by \citet{2013ApJ...779..189F} (derived by taking into account 8-13\,mJy of potential contamination of the flux density measurement) we estimate a spectral index between $\alpha\approx-0{.}5\ldots-1{.}0$.
This is shallower than the spectral indices of known halos of $\alpha=-0{.}9\ldots-1{.}9$ \citep{2012A&ARv..20...54F}.
We shall note that flux densities in Abell\,2069 are only measured with large uncertainties due to the low surface brightness of the diffuse emission.
The area of the lowest contour of the GBT map of Abell\,2069 at 1.4\,GHz by \citet{2013ApJ...779..189F} (depicted in Fig.\,9 of their paper) is covered by $\sim$\,40\,beams in our radio maps with natural weighting (see Fig. \ref{fig:chandra}).
In our map the flux density uncertainty of such an extended source would be $\approx$\,19\,mJy.
However, most of the emission originates from the two cluster components A and B which is also in agreement with the GBT map by \citet{2013ApJ...779..189F}.
Assuming a spectral index of $\alpha_{346}^{1400}=-1$ we estimate a radio power of $P_{1{.}4\,\text{GHz}}=2\ldots3\cdot10^{23}$\,W\,Hz$^{-1}$ .

The dynamical state of the whole cluster appears to be rather complex.
The offset between the X-ray and radio peak of `HA' may indicate that the major merger in the main component A is still in an early or very late stage \citep{2009A&A...499..371G}.
The evolution of the radio halo morphology during a galaxy cluster merger has been shown by recent simulations by \citet{2013MNRAS.429.3564D}.
Moreover, given the low peculiar velocity, A and B are potentially undergoing a merger.
It is not known if B is still in its first approach or if it already passed the core of the main cluster.
If the latter is true it is possible that the sub-cluster's cool core has not been disrupted during the core passage \citep{2008MNRAS.391.1163P}.
Thus we can only speculate about the origin of the source `HB', which resides in the rather poor X-ray environment of B.
Only a few halos have been found so far in such poor environments \citep{2012A&ARv..20...54F}.
However, in this particular case the interaction of B with the much more massive component A may induce more energy -subsequently available for particle acceleration- into the intra-cluster medium of B than it would be possible by a merger activity of an isolated cluster of similar size as B.
Alternatively, a minor merger in B could have introduced turbulence in the volume of B.
Interestingly, \citet{2009ApJ...704.1349O} have found a cold front in the cluster component B.
\citet{2008ApJ...675L...9M} reported a spatial correlation between radio emission and their cold fronts for two clusters, RX\,J1720+26 and MS\,1455+22, which both possess a mini-halo.
For mini-halos it has been suggested that gas sloshing induces turbulence which eventually causes radio emission \citep{2013ApJ...762...78Z}.
Instead of a originating from turbulent electron acceleration the diffuse emission in B might be also caused by highly relativistic electrons generated by hadronic collisions of the cosmic-ray protons with the ambient thermal protons \citep{2014MNRAS.438..124Z}.
Spectral studies of `HB' over a large frequency range might help to discriminate between both scenarios \citep{2014arXiv1403.6743Z}.
Finally, it is also possible that `HB' is caused by fossil plasma ejected during a former active phase of an active galactic nuclei (AGN) in B.

In case the source is a halo this would only be the second discovery of a double-halo system \citep{2010A&A...509A..86M}, but the first one which is located within two components of a single galaxy cluster.

\section{Summary}
We present deep radio observations of the complex galaxy cluster Abell\,2069 carried out with the WSRT at 346\,MHz and a radio observation with the GMRT at 322\,MHz.
The cluster consists of a main component (A) with clear signs for a recent merger and a smaller companion nearby (B).
We have found extended emission in both components.
The emission in the main component is elongated roughly along the same axis as the X-ray surface brightness and has a LLS of $\approx$\,750\,kpc.
We classify the diffuse emission in the cluster's main component as a radio halo and estimate its lower limit flux density to be $25\pm9$\,mJy.

We discovered extended diffuse emission in the companion B as well. 
Its flux density amounts to $15\pm2$\,mJy and shows a LLS of the order of 50\,-\,100\,kpc.
With the available data the nature of the source remains uncertain.
Chandra observations have revealed a cold front in B.
Those fronts have been found for several clusters with radio mini-halos.
Alternatively, this diffuse emission might be caused by fossil plasma ejected by an AGN, or turbulence might be induced by the interaction between A and B.

Our observations reveal that Abell\,2069 shows, besides known intricated dynamics and a complex X-ray surface brightness, a richness of radio features including extended diffuse emission.
Further studies of this cluster may help to shed light on how the interaction during cluster mergers causes diffuse radio emission originating from the intra-cluster medium.

\begin{acknowledgements}
The authors thank the anonymous referee for useful comments and suggestions which have significantly improved the manuscript.
MH and UK acknowledge financial support by the DFG, in the framework of the DFG Forschergruppe 1254 `Magnetisation of Interstellar and Intergalactic Media: The Prospects of Low-Frequency Radio Observations'.
R.J.W. is supported by NASA through the Einstein Postdoctoral grant number PF2-130104 awarded by the Chandra X-ray Center, which is operated by the Smithsonian Astrophysical Observatory for NASA under contract NAS8-03060.
This research has made use of data obtained from the Chandra Data Archive and the Chandra Source Catalog, and software provided by the Chandra X-ray Center (CXC) in the application packages CIAO, ChIPS, and Sherpa. 
Funding for SDSS-III has been provided by the Alfred P. Sloan Foundation, the Participating Institutions, the National Science Foundation, and the U.S. Department of Energy Office of Science. 
SDSS-III is managed by the Astrophysical Research Consortium for the Participating Institutions of the SDSS-III Collaboration. 
\end{acknowledgements}

\bibliography{literatur}
\bibliographystyle{aa}

\end{document}